# Monitoring the evolution of dimensional accuracy and product properties in property-controlled forming processes

Sophie Charlotte Stebner[1]*, Juri Martschin[2], Bahman Arian[3], Stefan Dietrich[4], Martin Feistle[5], Sebastian Hütter[6], Rémi Lafarge[7], Robert Laue[8], Xinyang Li[9], Christopher Schulte[10], Daniel Spies[11], Ferdinand Thein[12], Frank Wendler[13], Malte Wrobel[14], Julian Rozo Vasquez[15], Michael Dölz[1], Sebastian Münstermann[1].

[1]Integrity of Materials and Structures, RWTH Aachen University, Aachen, Germany
[2]Institute of Forming Technology and Lightweight Components, Technische Universität Dortmund, Dortmund, Germany
[3]Chair of Forming and Machining Technology (LUF), Paderborn University, Paderborn, Germany
[4]Institute for Applied Materials - Material Science, KIT Karlsruhe, Karlsruhe, Germany
[5]Fraunhofer Institute for Casting, Composite and Processing Technology (IGCV), Garching, Germany
[6]Institute of Materials and Joining Technology, Otto-von-Guericke Universität, Magdeburg, Germany
[7]Chair of Forming and Machining Processes, Technische Universität Dresden, Dresden, Germany
[8]Institute for Machine Tools and Productions Processes, Professorship Virtual Production Engineering, Chemnitz University of Technology, Chemnitz, Germany
[9]Institute of Metal Forming (IBF), RWTH Aachen University, Aachen, Germany
[10]Institute of Automatic Control, RWTH Aachen University, Aachen, Germany
[11]Institute for Production Engineering and Forming Machines, Technische Universität Darmstadt, Darmstadt, Germany
[12]Institut für Geometrie und Praktische Mathematik, RWTH Aachen University, Aachen, Germany
[13]Professorship Measurement and Sensor Technology, Chemnitz University of Technology, Chemnitz, Germany
[14]Process Engineering Group, Institute for Mechanical Process Engineering and Mechanics, Karlsruhe Institute of Technology (KIT), Karlsruhe, Germany
[15]TU Dortmund University, Chair of Materials Test Engineering (WPT), Dortmund, Germany
*corresponding author sophie.stebner@iehk.rwth-aachen.de

**Abstract**

As recent trends in manufacturing engineering disciplines show a clear development in the sustainable as well as economically efficient design of forming processes, monitoring techniques have been gaining in relevance. In terms of monitoring of product properties, most processes are currently open-loop controlled, entailing that the microstructure evolution, which determines the final product properties, is not considered. However, a closed-loop control that can adjust and manipulate the process actuators according to the required product properties of the component will lead to a considerable increase in efficiency of the processes regarding resources and will decrease postproduction of the component. For most forming processes, one set of component dimensions will result in a certain set of product properties. However, to successfully establish closed-loop property controls for the processes, a systematic understanding of the reciprocity of the dimensions after forming and final product properties must be established. This work investigates the evolution of dimensional accuracy as well as product properties for a series of forming processes that utilize different degrees of freedom for process control.

## 1. Introduction

An increased effort on monitoring metal forming processes is observed recently (Awasthi et al., 2021; Biehl et al., 2010; Jayakumar et al., 2005; Kumar and Das, 2022; Ralph and Martin, 2020). As product tolerances and production standards become ever stricter within the context of sustainability as well as resource efficiency, forming processes not only need to be open-loop but rather closed-loop controllable (Awasthi et al., 2021; Bambach et al., 2022). To a large extent, forming processes already rely on closed-loop controls for the inline adjustment of the processes, however, control of the components product properties is primarily open-loop (Hao and Duncan, 2011). This particularly means that the microstructure evolution (e.g. texture, dislocation density) during the forming process, which dictates the product properties (e.g. strength and stiffness) of the component, is currently largely disregarded.

This causes various disadvantages. A forming process, that does not take property fluctuations in the workpiece into account, will eventually lead to time intensive postproduction of the component (meaning downstream process steps that only arise do to the mechanical properties not meeting the requirements) and/or considerable amounts of scrap (Allwood et al., 2016; Polyblank et al., 2014). Implementing closed-loop controls into the processes that get feedback on the microstructure evolution (thus product properties) will, hence, enable the forming processes to be designed more efficiently as well as sustainably.

For a schematic overview of the problem as well as the proposed solution to establish a closed-loop control based on product properties, see **Figure 1.**

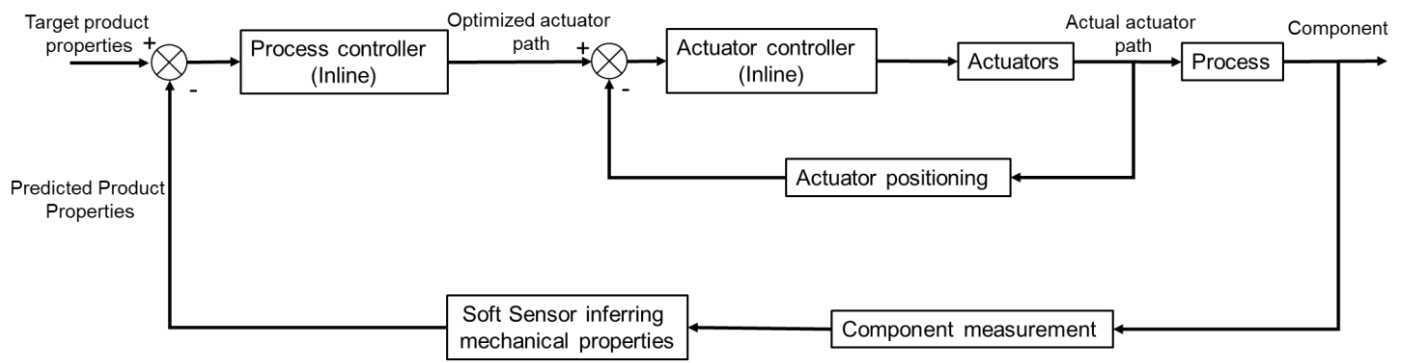

*Figure 1: Schematic depiction of proposed closed-loop control based on product properties adapted from (Polyblank et al., 2014)*

However, this endeavor harbors several challenges. First and foremost, it must be established whether and how sensors and actuators (in this case meaning all parameters in the process responsible in influencing the system, including forming tools but not limited to them) can be positioned in the processes, so that a microstructure evolution, hence product properties, may be detected and optimally manipulated. In most forming processes, actuators will have a highly non-linear effect on the product properties such as the dimensional accuracy and mechanical properties, making the model definition a complicated task, prone to many uncertainties (Allwood et al., 2009; Allwood et al., 2016). Regarding the position of the sensors, it is often difficult to measure the target values directly in the points of interest, meaning spatial and time-dependent analytical relationships between the process parameters and the product properties must be derived (Allwood et al., 2016). Furthermore, the sensor selection is limited to non- and semi-destructive testing technologies.

To overcome these challenges, one primary and initial step must be taken – establishing a systematic understanding of the interaction of the target values, i.e. the product properties and dimensional accuracy. This is especially challenging, yet essential for introducing a property-based closed-loop control in forming processes. This is mainly due to the fact that usually forming processes can form defined dimensions of a workpiece that are strictly tied to a set of product properties. Thus, it must be investigated if and how the evolution of properties can be decoupled from the evolution of dimensional accuracy.

To detect the microstructure evolution (consequently the product properties) during the forming process, a suitable sensor and/ or measurement technique must be identified. As the workpieces will need to be ready for application non- or at least semi-destructive sensors will need to be used. Non-destructive testing refers to the examination of materials and/ or components to determine their fitness for services without the degradation of the wanted properties. Non-destructive testing methods can be subdivided broadly into five categories – ultrasonic, electric, penetrant, electrical as well as radiographic (Halmshaw, 1991; Hinsley, 1959; Hull and John, 1988). Semi-destructive testing methods refer to methods causing small cavities in the component, however having low impact on it. Methods referred to as semi-destructive include penetration methods such as ultrasonic contact impedance (UCI-) hardness tests or pull-off methods (Jaskowska-Lemańska and Przesmycka, 2020; Jaskowska-Lemańska and Sagan, 2019). Depending upon the tested materials and materials characteristics, a suitable non-destructive or semi-destructive testing method must be established for use.

Furthermore, the utilized sensors' signals must be able to detect the microstructure evolution that needs to be correlated to the to be controlled product properties. In this study, processes are investigated, where particularly the development of microstructural phases influencing hardness, strain hardening, strength, and residual stresses of steel components during forming represent product properties of interest. Hence, especially electromagnetic sensors, such as magnetic Barkhausen noise (MBN) and eddy current testing (ECT), offer a potentially useful inline measurement technique as they are particularly susceptible to a microstructure evolution. These experimental techniques are briefly discussed in the following:

- The testing method of (MBN) can be described as follows: When an alternating magnetic field is applied to a ferromagnetic material (as is the case for most steels), the magnetization of the ferromagnet will occur discontinuously as well as abruptly. These discontinuous changes occur due to the spontaneous movement of so-called domain walls in the ferromagnet as they overcome microstructural hindrances, so-called pinning sites, such as inclusions, dislocations, grain boundaries etc. These jumps lead to electrical noises, that are then detected by the measurement equipment. Variations in the pinning site intensity versus the actual domain wall motion is then the foundation of the MBN method for characterizing the microstructure evolution, as this will ultimately lead to effects on the MBN characteristics (Ghanei et al., 2014; Pérez Benitez et al., 2020; Wang et al., 2012). Current MBN testing equipment include but are not limited to the 3MA-II sensor technology developed by Fraunhofer Institute for Nondestructive Testing (Fraunhofer Institute for Nondestructive Testing IZFP), the Rollscan-Barkhausen noise signal analyzers from developer Stresstech (Stresstech Oy) as well as the QASS µmagnetic testing equipment from QASS developers (QASS GmbH).
- During ECT a coil produces an alternating magnetic field. When a conductive material is subjected to this magnetic field, so called eddy currents flowing through the material in closed loops will be produced. These eddy currents then entail a secondary electromagnetic field which counterposes the primary electromagnetic field from the coil, hence causing measurable changes in the electromagnetic field of the coil. If a tested material contains discontinuities in the microstructure, the flow pattern of the eddy current is disturbed which in turn also influences the electromagnetic field of the coil (Ghanei et al., 2014; Halmshaw, 1991; Hinsley, 1959; Hull and John, 1988). Hence, ECT also offers a suitable method for the detection of the microstructure evolution during forming. Current ECT equipment includes but is not limited to the EddyCation Professional system developed at University of Magdeburg (Methodisch-Diagnostisches Zentrum Werkstoffprüfung e.V) and Eddy-Current Vector sensor (working title) developed at TU Chemnitz.

As it is now established, which inline measuring equipment may be used, it is necessary to identify a method that allows the description of the relationship between the now detectable process-induced evolution of the measurand and the product property that shall be monitored. So-called soft sensors pose as such a method. The term soft sensor is a fusion of the words "software" and "sensor". They typically consist of an arrangement of at least one sensor or measurand respectively in combination with a mathematical modeling approach. This combination is then used to establish a relationship between the measurand and the sought variable (Becker and Krause, 2010; Fortuna et al., 2007; Jiang et al., 2021; Kadlec et al., 2009). Soft sensors have found their application in several industries and continue to gain popularity in all kinds of disciplines, such as robotics (Goh et al., 2022; Jaffer and Young, 2009; Zou et al., 2021), medicine and pharmaceuticals (Arpaia et al., 2021; Dahiya et al., 2023; Gyürkés et al., 2022; Hwang et al., 2022; Qiu et al., 2022) as well as various engineering disciplines as petrochemical process design (Morais et al., 2019), autonomous driving (Sekhar et al., 2022) or even estimation of cement fineness (Stanišić et al., 2015). Dependent upon the to be monitored parameters, the industrial process and thus the resulting system, a suitable modelling approach must be identified. Generally, soft sensors can be differentiated broadly into two branches – model-driven and data-driven soft sensors. Model-driven soft sensors, also called white box models, are usually based on the underlying physical and/ or chemical framework of the process, meaning they have complete phenomenological knowledge of the process (Fortuna et al., 2007; Jiang et al., 2021; Kadlec et al., 2009). However, when the a-priori knowledge of the process, thus system is insufficient, or the modelling is disproportionately complicated, data-driven soft sensors, also called black box models, become relevant. They rely purely on empirical observations of the system. In between these two categories, many combinations can be formulated, taking advantage of these two extrema.

Generally speaking, soft sensor design is made up of three stages: (1) data and information collection, (2) design and (3) implementation (Fortuna et al., 2007; Jiang et al., 2021; Kadlec et al., 2009). Thus, the quality of the data must be established, meaning the positioning of the sensors and actuators must be chosen in such a way, that the effects of missing data, outliers etc. are minimized. Secondly, the model type has to be identified. The model type is dependent upon the systems nature and can be summarized by three parameters particularly: the linearity, the dynamics and last but not least the degree of intelligence of the model. Lastly, the selected model needs to be implemented (Fortuna et al., 2007; Jiang et al., 2021; Kadlec et al., 2009)

In the authors' case, particularly for the first two design steps, it must be determined how the measurand evolution and thus the product properties are influenced by the forming processes. Meaning, for the studied forming processes, how the formed geometry that is usually tied to a defined set of product properties may be decoupled from the evolution of relevant properties, as there is a lack in systematic understanding of decoupling strategies in different forming processes. Creating this understanding, it will ultimately allow the manipulation of the product properties during forming using different degrees of freedom (DOF), however still leading to the targeted dimensional accuracy of the workpiece.

Hence in **Chapter 2**, this paper will introduce and discuss options for the decoupling of dimensional accuracy and product property evolution during the forming processes. In **Chapter 3**, an overview on the anticipated decoupling strategies for selected forming processes will be given. **Chapter 4** consists of the detailed case studies of each selected forming process and their respective decoupling strategy. Finally, **Chapter 5** will conclude this study.

## 2. Strategies for decoupling of geometry and property evolution during forming

Product properties determine the application of the workpiece. Hence, it is important to first and foremost know the product properties of a component and, furthermore, to be able to influence them according to the required characteristics. During forming, the semi-product undergoes plastic deformation, leading to a permanent change not only in shape but also in the product properties. To understand the product property evolution during forming, the microstructure evolution that occurs must be investigated, as the microstructure dictates the product properties (Hornbogen and Warlimont, 2001). This holds true in particular for the strain hardening, which is mainly based on the increased dislocation density resulting from the forced plastic deformation. During forming, microstructural parameters as the density and distribution of dislocations, phase distribution as well as grain shapes and orientations, as well as damage in the material are affected, which in turn account for the final product properties as the strength, strain hardening behavior, hardness, and so on. This microstructure evolution during forming can be influenced, namely, by the temperature, the strain rate, and the state of stress.

The forming temperature, for example, has significant influences on the yield strength of the material. Generally, it can be said that the yield strength decreases with increasing forming temperature. For body-centered cubic metals, the strain hardening behavior of the material is not strongly influenced by an increase in temperature. However, in face-centered cubic metals it is due to the impeded movement of the dislocations with a decrease in temperature (Bleck, 2011; Doege and Behrens, 2010; Hornbogen and Warlimont, 2001). An increase in strain rate leads to an increase in yield strength and tensile strength, whereas uniform elongation decreases. This holds true for strain rates below $\approx 10^{-2}$ 1/s. Whenever, this critical value is exceeded, the heat exchange between the workpiece and surrounding is not sufficient. This leads to a heating of the component, which in turn again leads to a decrease in yield and tensile strength (Bleck, 2011; Doege and Behrens, 2010; Hornbogen and Warlimont, 2001). Furthermore, as a result of the forming process, the stress state, state of damage as well as the texture in the component changes. In terms of stress state, hydrostatic stresses must be mentioned as they do not particularly influence the level of plasticity, though may be used to suppress or foster damage evolution. However, they do lead to a change in microstructure in distorting the lattice, hence, favoring diffusion (Gao et al., 2011).

This shows that the macroscopic product properties are dictated by microscopic occurrences that are strongly dependent upon external factors. Hence, the reciprocity of process, microstructure and product properties needs to be fully understood to not only open-loop control processes but rather to closed-loop control them. Several studies regarding the microstructure evolution of steels during various processes have already been conducted. Just to name a few, the authors in (Bontcheva and Petzov, 2003) introduce a mechanical model for the numerical simulation of die forging, within which the microstructure evolution is coupled to thermo-mechanical phenomena. Reimann et al. (Reimann et al., 2019) use machine learning algorithms to numerically model the mechanical response of various microstructures subjected to loads based on representative volume elements and crystal plasticity modeling. Yang et al. (Yang et al., 2010) propose a model simulating microstructure evolution during induction heating for austenitization of steels based on cellular automata. The authors in (Salehiyan et al., 2020) study the microstructure evolution during deformation of a quenching and partitioning steel (QP980) and Vajragupta et al. (Vajragupta et al., 2020) present a micromechanical model for the mechanical behavior of a DP600 steel during sheet metal forming.

Investigations utilizing these approaches to systematically influence the interaction of the macroscopic product properties and the forming dimensions for the implementation of closed-loop controls of processes are scarce to non-existent. And it is precisely in this research gap that this study aims to contribute. Ultimately, this newly established understanding will allow not only to describe the interaction between processes, microstructures and product properties. Furthermore, it allows the extension onto the component performance, meaning the loading capacity of a component in the relevant load case for the design relative to the weight of the component.

# 3. Theory and Modelling of Selected Forming Processes

In the previous two chapters, it has been established that the described phenomena are occurring on different scales and need to be fully understood. However, starting on the smallest scale, detailed descriptions of the microstructure evolution which determine the product properties during forming used for the implementation of inline closed-loop property controls are still to be proposed. This is where this study aims to make an impact. In giving a detailed description for selected processes, see **Chapter 4**, for methods to influence the interaction of dimensional accuracy and product properties, a new engineering foundation for closed-loop property controls is laid. The processes outlined are:

(1) Skin-pass rolling: Skin-pass rolling is a process used to set the desired surface finish of strips. It uses textured rolls to imprint surface asperities on strips with minimal thickness reduction. Conventional methods use a single roll pass and adjust the roughness transfer ratio by changing process parameters such as thickness reduction. However, this limits flexibility and the range of achievable product properties for the subsequent processes such as painting and sheet metal forming. In this process, the roughness parameters of the strip $R_a$ and $RP_c$ are the surface properties of interest, as they influence the friction coefficient, the paint adhesion as well as paint appearance in the following processes. These surface properties are independently controlled by utilizing two roll stands and the corresponding strip tensions and are measured by integrating a contactless roughness sensor.

(2) Flow-forming: Flow-forming is an incremental forming process used to produce rotationally symmetrical hollow components with high dimensional accuracy, for example, in drive technology or in the aerospace industry. The process is characterized by high cost-effectiveness compared to metal-cutting processes. In addition, the components have a very high surface quality and improved mechanical properties due to the process-related high strain hardening.

(3) Open-die forging process: According to DIN 8583, the open-die forging process belongs to the free-forming processes (Deutsches Institut für Normung e.V.). In this process, the workpiece can be moved and formed freely between two open dies. This method of open-die forging, therefore, offers an ideal basis for locally influencing the microstructure development during the forming process. This is intended to improve the creep and fatigue properties. For the microstructure transformation, the semi-finished product must be heated locally. Therefore, an appropriate tool has to be designed that allows conductive heating during the open-die forging process. Targeted control can be used to decouple the development of dimensional stability and product properties by controlling the microstructure evolution.

(4) Thermomechanical Tangential Profile Ring Rolling: Tangential profile ring rolling is a highly efficient process for manufacturing ring-shaped parts such as bearings. By applying specific process parameters with regard to temperature and rate of deformation, it is possible to affect the microstructure, strength, and surface hardness of the final product during the rolling process, without requiring secondary heat treatment. Due to the complex interactions of initial material state, material physics and machine, a closed-loop control is needed to achieve the desired final state.

(5) Punch-hole-rolling: The punch-hole-rolling process is an incremental sheet forming process creating a rimmed hole with a cylindrical interior surface in the sheet plane. This shape can be applied for example as a fixture for bearings in machine parts of logistic systems or automotive drivetrain components. To keep a maximum of flexibility with different bearing types and sizes the processing steps of punching and rolling must be tuned to reach required geometric tolerances, microstructures and retained strength and ductility of the part. A sufficient balance and control of these properties is essential to guarantee reproducible and durable metal sheet products from the punch-hole-rolling process.

(6) Reverse Flow-forming: Reverse flow-forming is an incremental process that offers flexibility and efficiency. Through an intended wall thickness reduction, it allows the production of tubular parts with excellent shape, dimensional accuracy, and surface qualities that meet the stringent requirements of industries such as aerospace. Applications include drive shafts for jet engines and helicopters. A unique aspect of reverse flow-forming is that the material flows in the opposite direction of the roller tool movement. This process is difficult to predict in terms of stress and strain distribution and is affected by many factors such that adjusting local workpiece properties, as microstructure profile, remains a challenge.

(7) <u>Freeform bending:</u> Freeform bending with movable die is a process allowing the bending of complex geometries without having to change the bending tool. The process finds its application in the automotive and aerospace industry as well as in energy and chemical production. All these industries, however, are facing problems with post processing steps of the freeform bent parts. Thus, product parameters of interest are residual stresses, strength as well as the induced plasticity during bending as these particularly dictate further processing steps as well as service application of the component.

(8) <u>Press hardening:</u> Multi-stage hot sheet metal forming allows the forming of complex shaped press-hardened components at high stroke rates. Initially, rectangular blanks are heated to the austenitization temperature $T_\gamma$ by means of induction heating, whereafter, a temperature profile is set using resistance heating and cooling by compressed air from flat fan nozzles. In the first forming stage, the blank is stretch-formed into a hat shaped profile followed by die bending, where one sidewall of the hat shaped profile is bent. However, product properties that are influenced by thermal and mechanical influences, such as hardness and sheet thickness, are difficult to predict, thus limiting the application of the process.

As types of forming processes are manifold, they can further be categorized regarding their forming stages, whether they are hot formed or cold formed and what kind of material is used. The authors propose the following categorization, see **Figure 2**, for the forming processes based on which general insights for an implementation of a closed-loop property control may be gleaned in **Chapter 5**. The numbers in the pictograms correspond to the numbering above.

| *Forming Stages* | *Forming Temperature* | *Materials* |
|---|---|---|
| **Single-Stage Forming Process:** 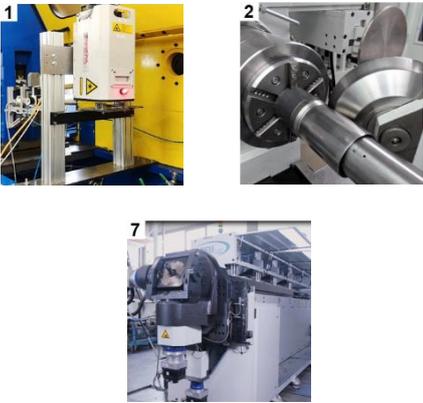 | **Cold Forming:** 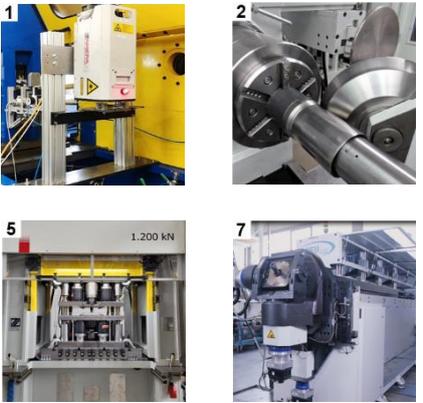 | **Ferromagnets:** 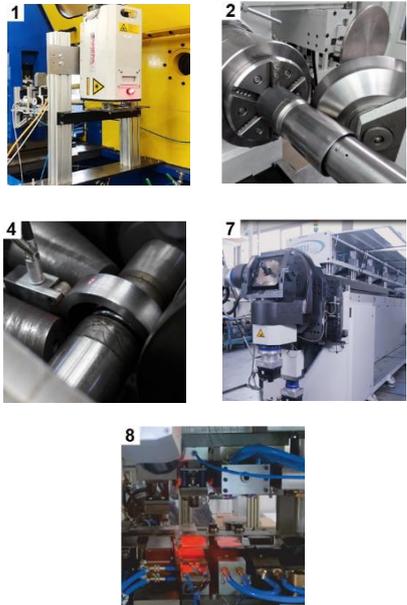 |
| **Multi-Stage Forming Process:** 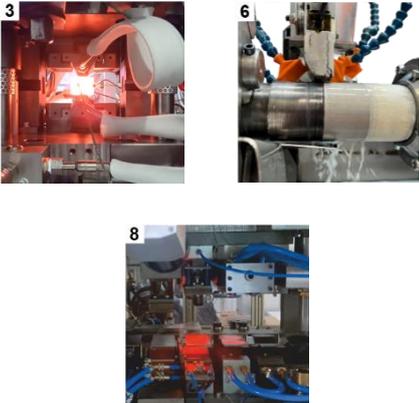 | **Hot Forming:** 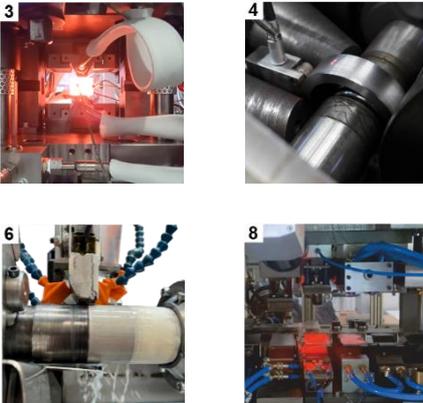 | **Paramagnets:** 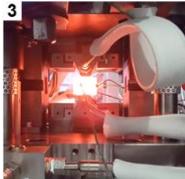 |
| **Cyclical Forming Processes:** 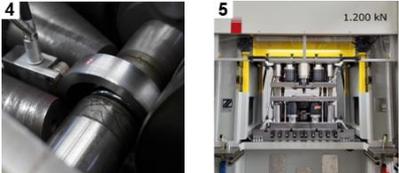 | | **Metastable Paramagnets:** 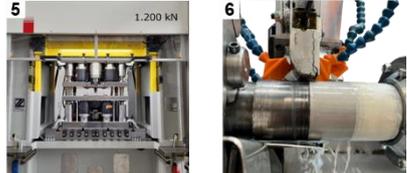 |

Figure 2: Categorization of forming processes according to forming stages, temperature and Materials investigated

## 4. Case Studies and Results
### 4.1 Skin pass rolling (Single-stage, Cold, Ferromagnetic Material)

Skin-pass rolling is a rolling process that is typically used to achieve a specific surface finish on strips after they have been subjected to conventional cold rolling. This process uses rolls with textured surfaces to imprint a certain roughness on the strips, resulting in a minor reduction of their thickness. The textured surfaces on the rolls are created using various techniques such as electrical-discharge texturing (EDT), shot-blast texturing (SBT), or laser texturing (LT), and can be either stochastically or deterministically distributed. The surface roughness of the resulting sheets, which is characterized by parameters such as the average roughness $R_a$ and peak number $RP_c$, plays a crucial role in determining the product properties of the sheets, including their tribological properties in the subsequent forming process (Emmens) paint adhesion, and paint appearance after painting (Scheers et al., 1998). The effectiveness of the surface texturing process is determined by the roughness transfer ratio, which compares the outgoing roughness on the strips to the roughness on the rolls.

Conventionally, only one rolling pass is employed in skin-pass rolling. To achieve the desired outgoing roughness on strip surfaces, the roughness transfer ratio is to be adjusted by changing the process parameters such as thickness reduction. To save the expense of manufacturing new textured rolls and reduce labor costs for roll changing, the roughness of the work rolls cannot be influenced. Therefore, in conventional skin-pass rolling, thickness reduction is always coupled with the final surface roughness of strips, which reduces the flexibility of the process. Moreover, the $R_a$ and $RP_c$ cannot be controlled separately, thus, the achievable range of product properties is restricted as well.

In this study, a novel approach for the skin-pass rolling process is presented. The system features a two-stand rolling configuration, comprising a pair of untextured and textured (EDT) rolls respectively. The strip tensions for these two stands are regulated by three dancer rollers positioned before, between, and after the rolling stands, as illustrated in **Figure 3**. In this system, the maximum rolling force is limited to 50 kN; the tension adjustment ranges of the three dancers are [50, 600] N, [50, 1500] N, and [50, 1500] N, respectively. Additionally, rotary encoders and thickness sensors are employed to monitor the position and thickness of the strip, and a contactless roughness sensor is installed at the end to facilitate real-time optimization of the surface texturing model ($R_a$ and $RP_c$). By manipulating the strip tensions (Schulte et al., 2021b) and controlling the distribution of the total height reductions over the two passes, this system offers greater control over the final product properties of the strip. Furthermore, the use of the textured (EDT) rolls and strip tension control, allows to optimize the surface finish while keeping the thickness reduction to a minimum.

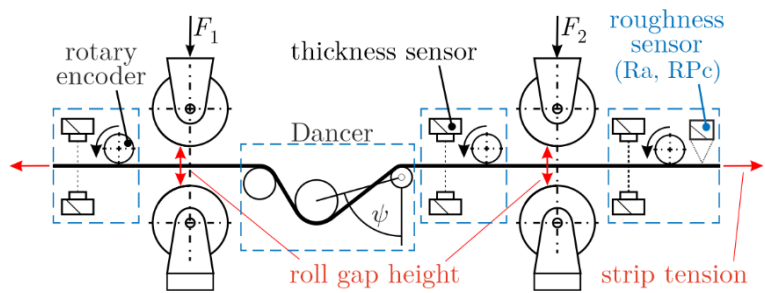

*Figure 3: Schematic depiction of skin pass rolling.*

To model the rolling process and the associated geometry change and resulting properties, a total of three models were developed or implemented, capable of predicting the resulting strip thickness (Wehr et al., 2020), roll stand deflection (Schulte et al., 2021a), and strip roughness (Schulte et al., 2021b) at the exit of each individual rolling stand.

By incorporating both roll stands and adjusting the strip tension, the total height reduction can be divided into two passes utilizing distinct roll surfaces. This approach not only enhances control over the resulting strip roughness, but also enables manipulation of the stress profile in the roll gaps, thereby affecting the neutral point and contact stress level. As a result, the imprinting in the roll gap can be independently controlled for both strip geometry and surface finish.

A superordinate adaptive control system is implemented to determine the optimal strip tensions and desired strip thickness for the first stand based on the desired outgoing strip thickness and roughness. To achieve this, a nonlinear model of the rolling mill is used to create an optimal control problem (OCP), which balances the deviation of the controlled variables and the change of the manipulated variables.

To solve this OCP in real-time, it is formulated as a sequential quadratic problem and iteratively solved using the qpOASES solver. The calculated references for strip tensions are then applied to the rolling mill, while the desired thicknesses for the two rolling stands are supplied to a subordinate strip thickness control scheme.

The decoupling of the strip thickness $h_2$ and strip roughness $Ra_2$ could be demonstrated in the following experiment, see **Figure 4** (Schulte et al., 2023). Here, the total height reduction was divided between two rolling stands $s_1$ and $s_2$ in order to control the desired roughness $Ra_{2,des}$.

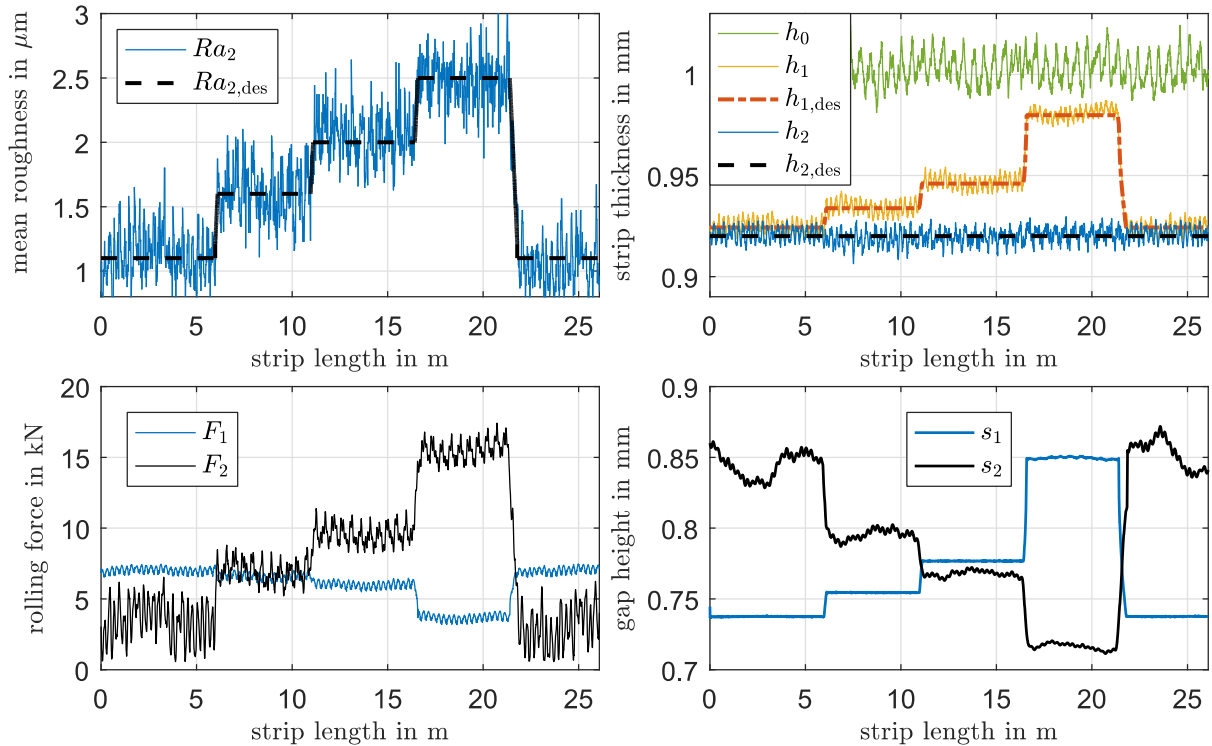

*Figure 4. Experimental results of an independent control of strip thickness ($h_2$) and roughness $Ra_2$, using only the two roll stands, compare (Schulte et al., 2023)*

### 4.2 Flow-Forming (Single-stage, Cold, Ferromagnetic Material)

Incremental forming processes (e.g. metal spinning, shear forming or flow-forming) are used to produce mostly rotationally symmetrical components from flat blanks or tubes. During the forming, the semi-finished product rotates together with a mandrel and is formed by the roller feed (**Figure 5**). The control of such processes is usually path based, i.e. it is not possible to react to disturbance variables such as variations in the geometry of the semi-finished product, variations in material properties, or process disturbances (friction, tool wear). The product property strain hardening is a positive result of the forming process, but are themselves only considered to a limited extent in the process design. The aim of this project is to take strain hardening into account in the design of incrementally manufactured components and to use it as a controlled variable during forming. For this purpose, strain hardening needs to be measured during the process. As there is no sensor that can measure the strain hardening during the forming process, a new sensor system has to be developed. This consists of several coils for determining the magnetic properties and a further temperature sensor. A soft sensor calculates the strain hardening based on the magnetic properties using a material dependent map. This is used as the basis for controlling the roller feed to achieve the desired strain hardening in the component. An advantage of the property-controlled flow-forming is that the forming machine itself can be used as the actuator, which means it can be used on almost any conventional flow-forming machine. In addition, the control system improves the process result independent from external influences. The new approach allows to produce components with graded strength, which can be used for crash components. It also makes it possible to achieve the optimum ratio between process time and strain hardening of the component during process design.

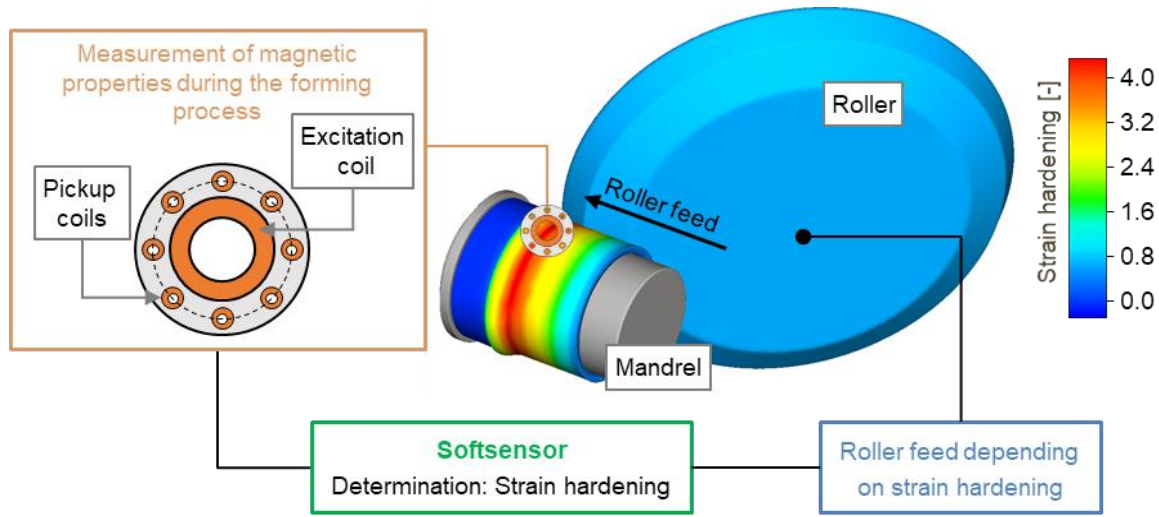

Figure 5: Approach of the property-controlled flow-forming.

In general, flow-forming produces a high plastic strain at a low roller feed rate and a lower plastic strain at a high roller feed rate. This is due to the different flow proportions resulting from the different displaced volumes displaced per mandrel revolution. This affects, for example, the pile-up volume of the material in front of the roller. In the case of a large pile-up volume, the stretching and subsequent compression of the material as it is formed under the roller to its final diameter results in greater plastic strain and thus greater strain hardening. The produced dimensional accuracy (e.g. final tube diameter) is therefore independent of the roller feed, as the final geometry is mainly determined by the gap between the roller and the mandrel. Roller feed can therefore be used as a DOF to control strain hardening, thus decoupling the evolution of dimensional accuracy and product property. In this way, the property-controlled process allows the production of parts with identical geometry but defined work hardness variations.

To adjust the roller feed rate during the process a real-time measurement of the plastic strain is required, independent of geometric variations like working distance and wall thickness. The determination of plastic strain and the resulting calculation of strain hardening by the flow curve within the soft sensor is based on the coevolution of mechanical and magnetic properties of ferromagnetic materials in the case of plastic strain. By enhancing the plastic strain and the resulting increase of defects the mechanical embrittlement coincides with the magnetic embrittlement, a reduction in the ability to adapt to external magnetic fields results. In addition to the effect of magnetic embrittlement, the induced stress in the material will change the ability to magnetize in different directions depending on the direction of the induced stress by the Villari effect (Joh et al., 2013). In this regard, a multi-sensor system (see **Figure 5**) has been designed to characterize the magnetic embrittlement in the form of the relative magnetic permeability and the directional variations in the form of the magnetic anisotropy in a single measurement. The multi-sensor system also has to quantify and compensate for geometric influences such as working distance, tilting of the sensor, and the curvature of the workpiece. For this reason, the central excitation coil is used for the method of inductance spectroscopy to independently quantify the working distance between the sensor and workpiece and the magnetic permeability to target magnetic embrittlement (Lu et al., 2018). The pickup coils use the excitation field of the sensor coil to quantify directional variations of the magnetic flux density on the circumference of the excitation coil allowing the separation of the magnetic anisotropy and effects due to the tilting of the sensor (Wendler et al., 2021). The directional properties of the magnetic anisotropy are used to evaluate contributions from the Villari effect and mechanical stress.

The sensor system does contain a soft sensor with an internal material model to use the information on the magnetic anisotropy and the changes of magnetic permeability to estimate the plastic strain as feedback for the control loop of the forming process (Laue et al., 2021).

### 4.3 Open-die forging process (Multi-stage, Hot, Paramagnetic Material)

Isothermal forging heats the tools to the workpiece temperature. This makes near-net-shaping feasible for materials with poor machinability. Advances in isothermal forging have recently made it possible to use turbine blades made of intermetallic titanium aluminides (TiAl) in serial production in commercial jet engines. However, there are currently many limitations to this technology can be found. The cast stock material consists of coarse lamellar colonies (**Figure 6 (a)**) and has highly anisotropic forming properties. For this reason, the preform has to be oversized. This is necessary to ensure that the degree of deformation is high enough to transform the cast structure. With current processes and tooling technology, the material utilization of turbine blades is often less than 10%. However, the closed-die forging process does not allow for a localized control input into the workpiece.

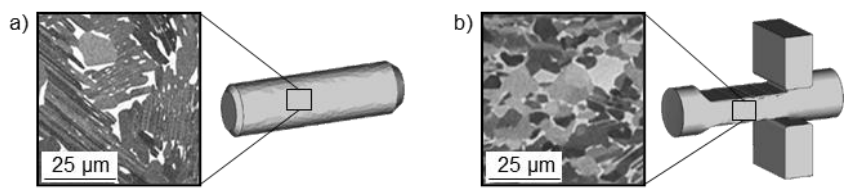

*Figure 6: (a) Initial Lamellar Microstructure; (b) Remodeled Globular Microstructure*

This research project aims to develop a multi-stroke, property-controlled open-die forging process in which the heating of the workpiece area to be formed takes place locally. In contrast to swaging, the open-die forging process allows the use of additional DOF for the closed-loop control of the property-determining globularization kinetics in the process. This is because the globularization kinetics of a TiAl alloy are subject to both material and process related disturbance variables. The knowledge and results obtained can be transferred to other materials and forging processes. The prediction of the microstructure development in the workpiece is based on level set formulations and neural networks, which act as a soft sensor.

The tests are carried out on a servo screw press of the type SHP-400 from Nidec SYS GmbH, Grafenau, Germany (Feistle et al.). The servo technology in conjunction with a machine control system adapted by Fraunhofer IGCV makes it possible to actively influence the course of the stroke curve $k$ and thus the ram stroke speed $v$ during the stroke. This represents a significant DOF in process control, since the prevailing strain rates $\dot{\varepsilon}$ in the material and the globularization kinetics can be influenced. The globularization kinetics are also influenced by the temperature fields present in the process. Consequently, the process temperature $T$, and thus indirectly the power $P$ of the source of the conductive heating unit, is considered as another DOF. The globalization of the material structure (**Figure 6 (b)**) depends on the locally prevailing degree of deformation $\varepsilon$. By integrating the feeder into the test setup (**Figure 7 (a)**) in combination with the variable mold closing height per stroke, the degree of deformation and the globularization kinetics can be actively influenced by additional DOF (**Figure 7 (b)**).

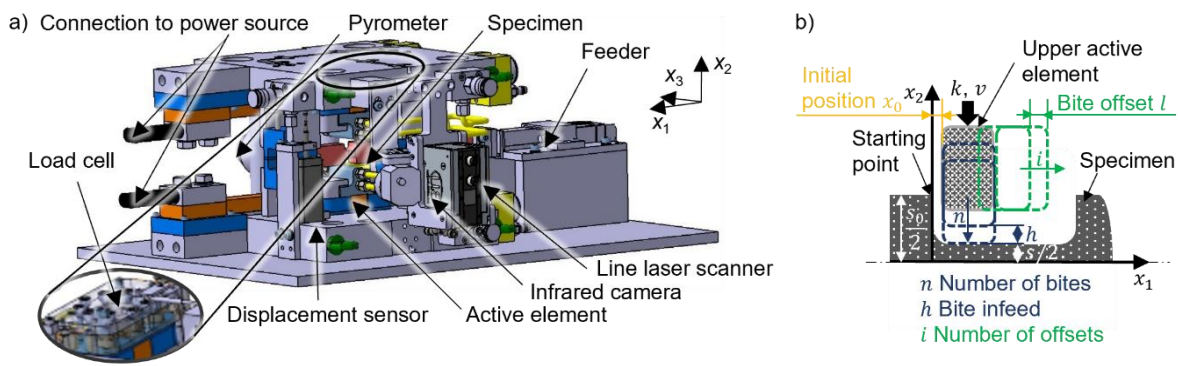

*Figure 7: (a) Schematic Draft of the Open-Die Forging Tool with Integrated Actuators and Sensors; (b) Degrees of Freedom of the forming process*

The variable mold closing height per stroke makes it possible to define the DOF $h$ of the height of the material to be formed, which is reflected in the stroke curve $k$. The number $n$ of blow sequences results from the desired residual material height $s$ and the height $h$ to be formed of the individual stroke as a function of the material start height $s_0$. The positioning of the specimen at the beginning of a stroke sequence is described by the DOF $x_0$. Between the strokes of a blow sequence, the specimen can be displaced by the feeder in $x_1$ direction. The displacement is described by the DOF $l$. The amount of displacement per stroke sequence is kept constant. The number of displacements in $x_1$ direction represents the DOF $i$. The geometry of the active elements of the tool can be considered as a further DOF, since it influences the degree of deformation as well as the material flow and thus the globularization kinetics in the material region to be formed. In this research project, the geometry of the active elements is not varied. The drive of the machine and the feeder are the initial actuators of the considered open-die forging process. The influence of the described DOF on the recrystallized material volume is visualized in **Figure 8** on the basis of a half-model of an FE simulation. For better comparability, the recrystallized areas are plotted against the sample geometry (bottom row of images, **Figure 8**). The residual material height $s$ of the formed area after the forging process is identical in both simulations. The material and damage model as well as the recrystallization of the microstructure according to the globularization kinetics were considered by a subroutine (Bambach et al., 2016; Imran et al; Imran et al., 2020; Imran and Bambach, 2018) in the forming simulation. Simulation validation was performed by comparing the globularized material volume from simulation and experiment.

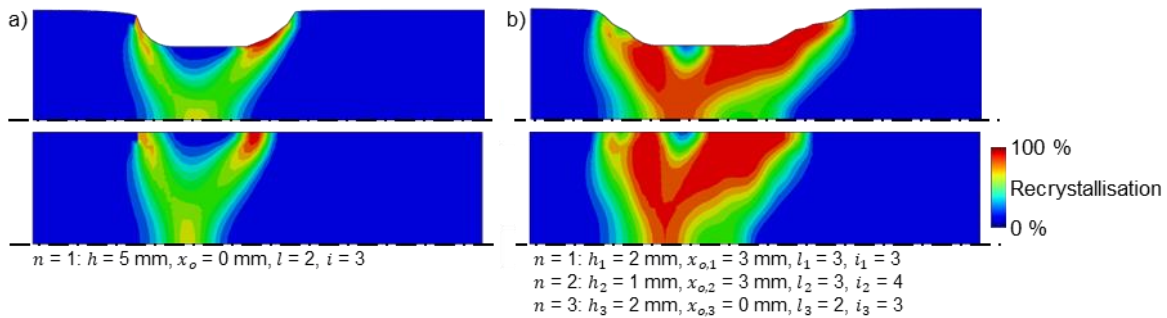

*Figure 8: Influence of the DOF $x_0$, $h$, $l$, $i$ at constant process temperature and velocities on the globularized material volume*

In the forging process, the forging temperature is considered a central process parameter, but the forging temperature is rarely varied during the forging process, so the power source is considered a subordinate actuator.

In particular, the DOF of the impact sequence and the DOF of the forging temperature or the power of the heating unit allow the material properties to be actively influenced. The boundary condition residual material height $s$ and the degree of freedom of the geometry of the active tool elements primarily influence the geometry of the component to be produced.

To control the manufacturing process, process data must be collected. In this research project, the previously discussed DOF are used to characterize the globularized microstructure during forming. The relevant process parameters are recorded by specially selected sensors. The displacement of the active element involved in the forming process is measured by a displacement sensor. The recorded displacement signal can be used to derive the stroke curve and, in particular, the movement speed $v$ of the active element, the prevailing global strain rate $\dot{\varepsilon}$ and the residual component height $s$. The integrated line scanner allows the component geometry to be recorded along the $x_1$ axis of the test specimen in the area of the forming die. By combining the characteristics of the displacement sensor and the line scanner, the deformation $\varepsilon$ can be determined. The recorded characteristics of the pyrometer are used to control the power $P$ of the heating unit. The three sensors discussed can be considered primary sensors for controlling an open-die forging process. Additional information is provided by the load cell and the infrared camera integrated in the test setup. The data from the load cell is used for process monitoring to prevent overstressing of the system or the forming tool. The integration of the infrared camera makes it possible to evaluate a wider range of materials in terms of the prevailing temperature distribution and the historical temperature profile.

The coupling of actuators, sensors, machine control, and the implemented controller is shown in **Figure 9**. The figure also shows the integration of the soft sensor for predicting the globularization kinetics. A detailed description of the implementation and the interfaces has already been shown in (Feistle et al.).

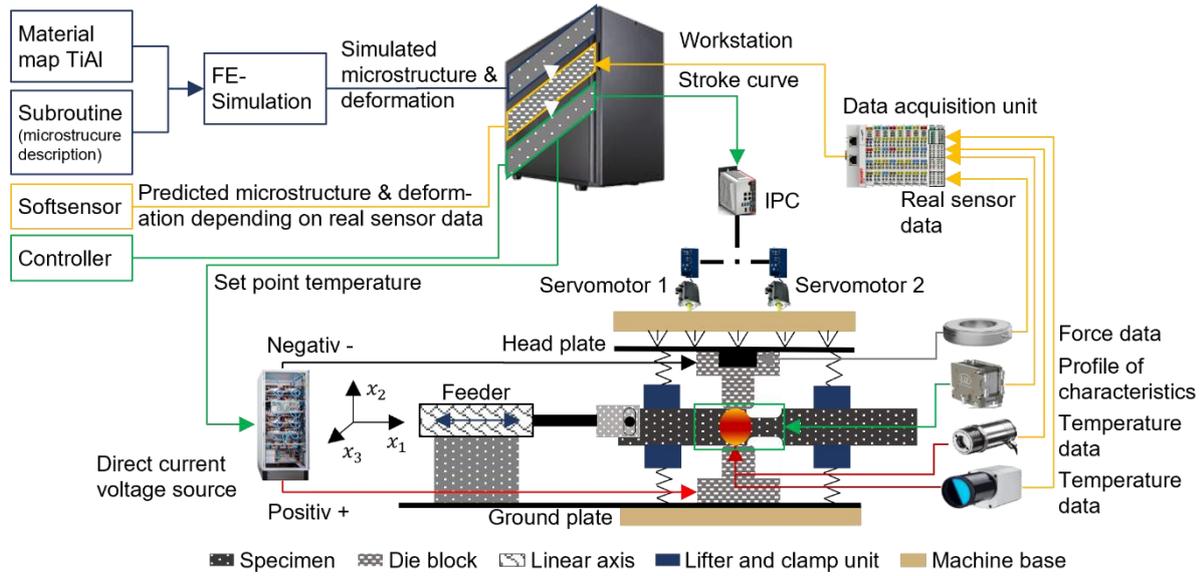

*Figure 9: Process Setup and Control-Loop in Open-die Forging for Local Control of Globularization Kinetics and Geometry*

The soft sensor is based on a Conditional Variational Autoencoder, which is an extension of a Variational Autoencoder. Through training, the generative model learns a latent representation of the input data (Kingma and Welling, 2013). A Variational Autoencoder consists of an encoder network and a decoder network. The encoder network maps input data to a distribution in latent space, and the decoder network maps points in latent space back to data space. The goal of a Variational Autoencoder is to maximize the lower bound of the log-likelihood of the data. The Conditional Variational Autoencoder conditions the encoder and decoder on additional information that the Variational Autoencoder does not. The conditioning is typically done by concatenating the information to the input of the encoder and decoder. This allows the Conditional Variational Autoencoder to generate samples with specific attributes, making it useful for tasks such as image generation and information retrieval (Sohn et al.). The neural network training data is based on FE forging simulations performed. The network can be used to estimate not only the globularization kinetics, but also the deformation of the test specimen and the prevailing degree of deformation $\varepsilon$. The recrystallized material volume in the half-model, shown in **Figure 10 (a)**, of a forging sequence performed is superimposed in the next step with the input vector of the process data of the upcoming forging cycle and thus serves as an input variable for the Conditional Variational Autoencoder. **Figure 10 (b)** shows the simulated reference state of the recrystallized material volume of the final forging process. **Figure 10 (c)** shows the estimated distribution of the recrystallized volume by applying the Conditional Variational Autoencoder. A comparison between the FE model and the Conditional Variational Autoencoder shows that the deviation is in the single digit percentage range.

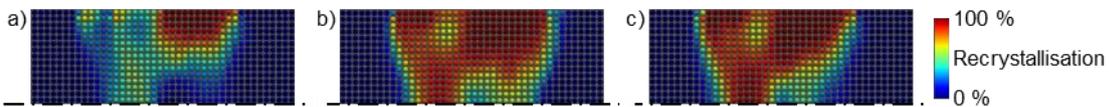

*Figure 10: (a) Input Image of the Degree of Recrystallisation of the Previous Stroke; (b) Calculated Degree of Recrystallisation by FE-Simulation Mode l; (c) Determined Degree of Recrystallisation by Conditional Variational Autoencoder*

The mathematical modeling of the underlying dynamics is done via a level set $\phi(t,x)$. This level set describes the region of the already transformed microstructure with the desired globularization $X$. The evolution of the level set is described by a Hamilton-Jacobi equation (**Formula 1**)

$$\frac{\partial \phi}{\partial t} + H(x, \nabla\phi) = 0, \phi(0,x) = \phi_0(x), \phi(t,x_0) = f(t,x_0) \qquad \text{Formula 1}$$

By a suitable reformulation, the deviation from a desired reference state can be described by a system of hyperbolic partial differential equations. For this system, a control rule was determined in (Herty and Thein, 2022), which minimizes a disturbance exponentially in time.

### 4.4 Thermomechanical Tangential Profile Ring Rolling (Cyclical, Hot, Ferromagnetic Material)

Ring rolling is a forming process for the fast and inexpensive production of ring-shaped parts. In particular, tangential profile ring rolling (TPRR) enables the production of bearing rings or similar parts with near net-shape geometry and dimensions and advantageous material properties due to a high degree of strain hardening at a high cadence. Further control of material properties can target specific phase compositions, grain sizes and/or additional boundary layer properties. To achieve this goal, the required thermomechanical treatment must be carried out during the forming operation, without disturbing the final geometric shape (Brosius et al.). This can only be achieved by decoupling the influence of various actuators with regard to their effect on the net shape and on the microstructure evolution and then employing a closed-loop predictive control system in the process. In particular, as the part's shape is generated by the tool geometry, the final geometry results from the positioning of the rolls at the end of the process. The path toward this final alignment gives an almost arbitrary DOF that can be used as a method to influence the material properties. Both the rolling force and the amount of forming per part rotation can be controlled separately within a fairly large process window and without additional actuators already result in different microstructural results due to partial dynamic recrystallization (Lafarge et al., 2023). An even larger influence can be reached by also controlling the forming temperature during the process. To avoid unnecessary complexity and as no advantages could be gained from a re-heating of the part, this control is only done via active cooling of the part with compressed air.

The overall forming process can then be controlled by cascading closed loop control responsible for the different aspects of the aspect. This design also offers a lower complexity, as the property control loop can use separately controlled physical conditions such as a temperature value as its actuating variable instead of physical actuator inputs. This was validated for the cooling rate control loop.

For this model-predictive closed-loop control, multiple sensors are required to evaluate the current process conditions. Apart from internal information such as the current advance, direct measurement of the force and surface temperature is performed. Due to variations of the material properties, additionally the microstructure needs to be evaluated in a non-destructive way during the forming process. In this application, this is done by an ECT system monitoring the outside of the rotating ring. This setup requires an additional step, as only near-surface information can be gained that way. This information is integrated into a state model that is constantly re-aligned to the measurement obtained by ECT sensing. This in-process model is based on a linearized characteristic model that is derived from a large number of FEM simulations of various process conditions, including those that cannot be (safely) reached in practical experiments. For the simulation of geometry and microstructural properties, the use of a pseudo plain strain FE model enables to determine strain and temperature. This data can then be used for the computation of microstructure and hardness.

For experimental validation of these concepts, a ring rolling machine 1986 UPWS 31,5.2 ring rolling machine from VEB Werkzeugmaschinenfabrik Bad Düben (Ficker et al., 2005) was heavily modifed. This machine has been altered in several ways to allow for individual control of: rolling force, rolling rate and an additional cooling by several compressed air nozzles allowing for air flow rates between 0 and 500 l/min (see **Figure 11(a)**) for a schematic view of the machine kinematics). All of these can be controlled as functions of time. The forming tools consist of a main roll which pushes the workpiece towards a free-rolling mandrel. Main roll and mandrel together contain the desired ring profile. Additionally, sensors were added to measure tool displacement, rolling force, ring growth, ring surface temperature and microstructure evolution using a high-temperature ECT system.

In this example, ring blanks were manufactured out of 16MnCr5 (1.7131, AISI 5115) tube material which was then austenitized at 900°C for 30 minutes. Rings were then transferred to the rolling machine and formed using various parameter sets.

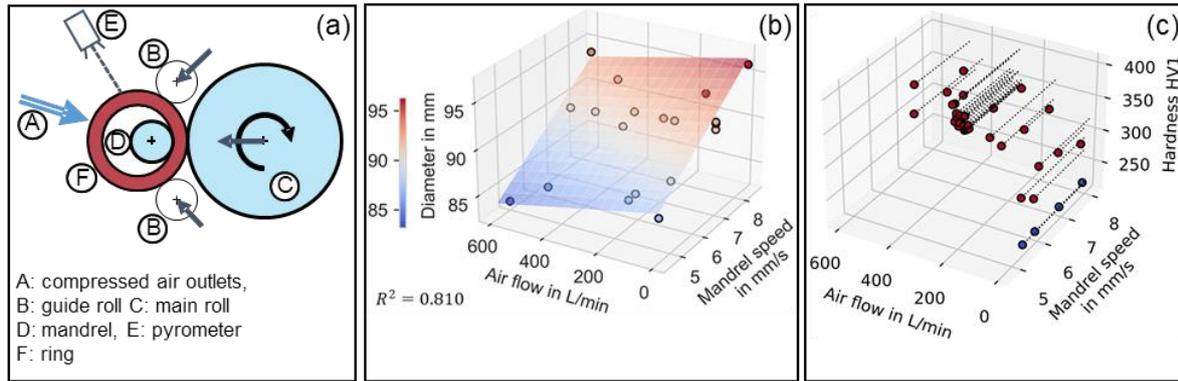

*Figure 11: (a) Ring rolling machine kinematics and actuators. (b) influence of air flow (corresponding to cooling rate) and mandrel speed on resulting dimensions (from (Kumar and Das, 2022)). (c) influence of the same parameters on hardness achieved.*

It is well-known that the final ring diameter is influenced by the main roll advance per rotation (Rolf and Perter, 1984). Additionally, **Figure 11(b)** shows that without further control, the diameter is also dependent on the cooling employed, chiefly because the flow stress is directly influenced by temperature (Lafarge et al.). The influence of those parameters on the achievable hardness is shown in **Figure 11(c)**. It can be seen that the desired product properties can be achieved over a range of parameters that leads to varying dimensional accuracy, but crucially these are not directly correlated. Since following a specific cooling temperature profile is key to achieving the desired microstructural changes (Hütter et al., 2021), it is clear that it is indeed not possible to simply follow a prescribed forming profile, even with precise control. Instead, to decouple the generation of dimensionally accurate parts from the product properties, model-predictive control (Lafarge et al., 2021) of the actuators as proposed above is required.

### 4.5 Punch-Hole-Rolling (Cyclical, Cold, Metastable Paramagnetic Material)

The punch-hole-rolling process (developed by the Institute for Production Engineering and Forming Machines (PtU) in cooperation with the Institute for Applied Materials (IAM-WK)) is an incremental forming process, which deforms sheet metal, however, shows strong similarities to bulk forming operations. In contrast to ordinary sheet forming processes, the sheet thickness is changed significantly, creating a collar on both sides of the sheet starting from a preformed hole $r_0$. The process can be used, for example, as a bearing seat for electric motors and offers the advantage that the bearing seat sits within the sheet plane and thus offers improved stability in the housing. A key product property is hardness, which improves the wear properties of the tribologically loaded bearing seat.

The collar is formed by a roller which is inserted into the hole at the start of the process and is then moved in a spiral pattern, simultaneously increasing the diameter of the hole and the height of the collar in dependence of the desired final radius $r_f$. The variable geometrical product properties are the diameter of the hole and the height of the collar. The variable process parameters are the radius of the starting hole $r_0$, the radial position of the roller $r(t)$, its radial feed rate $\dot{r}(t)$, and its angular velocity $\omega(t)$, see

**Figure 12 (a).** In order to be able to realize a closed-loop control of the forming progress in a way that allows the mutual variation of dimensional accuracy of the collar as well as the product property (e.g. hardness), a MBN sensor is used to indirectly measure the hardness in a non-destructive, rapid way. This measurement technique is currently implemented in the forming tool with a magnetization frequency of 125 Hz and a magnetization voltage of 9 V for in operando contactless measurements (sensor to workpiece distance 50 µm).

Previous tests show that the radial feed rate $\dot{r}(t)$ has a significant influence on the collar height and the martensite fraction, whilst changing the rotational speed only significantly affects the martensite fraction. Therefore, a decoupling of the evolution of dimensional accuracy and product properties respectively hardness is feasible. In recent investigations, tests were carried out by means of punch-hole-rolling 3 mm thick sheets of TRIP (transformation-induced plasticity) steels 1.4301 and 1.4404 in a semi-factorial variation of the individual process parameters ($r_0$, $\omega(t)$, $\dot{r}(t)$). Regarding the TRIP-Steels, it can be stated that the strain rate-dependent martensite formation is an advantageous property for process control (see **Figure 12 (b)**). It has already been shown in recent results that the martensite content depends on the one hand on the degree of forming and on the other hand on the strain rate (Fonstein, 2015), as can be modelled by the Olsen-Cohen model (Olson and Cohen, 1975a). Therefore, the martensite fraction can be controlled solely by varying the deformation speed at constant overall plastic deformation. In conclusion, this allows the hardness within the collar to be varied as the martensitic transformation is accompanied by an increase in hardness.

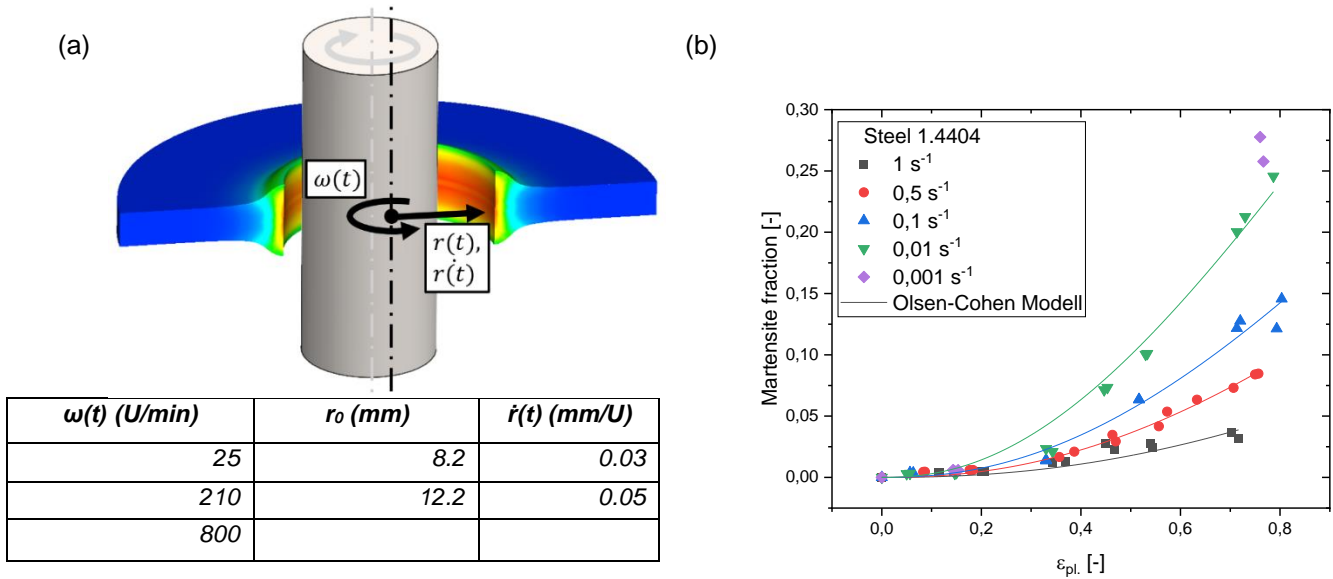

| $\omega(t)$ (U/min) | $r_0$ (mm) | $\dot{r}(t)$ (mm/U) |
|---|---|---|
| 25 | 8.2 | 0.03 |
| 210 | 12.2 | 0.05 |
| 800 | | |

*Figure 12: Schematic representation of the punch-hole-drilling process and its variable process parameters (a) and dependence of martensite fraction on strain rate and overall plastic strain for the steel 1.4404 (b).*

### 4.6 Reverse Flow-forming (Multi-Stage, Cold, Metastable Paramagnetic Material)

Due to the large number of disturbance variables, in reverse flow-forming is difficult to predict stress and strain distributions during the production process as well as the final product geometry, dimensions and properties (Mohebbi and Akbarzadeh, 2010; Runge, 1994). In conventional reverse flow-forming, the focus is put on the geometry and dimensional accuracy, specifically regarding the wall thickness and the elongation of the workpieces **Figure 13 (a)**. The present case study additionally aims to control local product properties. For this purpose, seamless tubes made of metastable austenitic steel such as AISI 304L (X2CrNi18-9 / 1.4307) are used as raw material. Since the austenitic phase is not in equilibrium, under specific temperature conditions phase transformation from metastable austenite into α'-martensite occurs during the deformation process (Nishiyama, 2014; Olson and Cohen, 1975b). This microstructure evolution entails changing of the magnetic and mechanical properties. On the one hand, the material transforms from paramagnetic austenite to ferromagnetic martensite. On the other hand, consequent to the increase of martensite in the microstructure, a strain-induced hardening process takes place (Knigge, 2015).

The reverse flow-forming process offers various mechanical DOF: the infeed and feed rate of the roller tool, the rotational speed of the mandrel, and the intended number of passes. Additionally, the local workpiece temperature in the forming zone has a significant impact on the final product properties that can be controlled by means of active cooling or heating, e,g, with cryogenic cooling or heat induction. In this case study a PLB 400 spinning machine from Leifeld Metal Spinning GmbH (Ahlen, Germany) with a drive power of 11 kW and a maximum achievable spindle speed of 950 rpm was used. Here, the hydraulically driven cross support is the main actuator moving the roller tool in axial $x$ and radial $z$ directions. The cross support is used to control the infeed depth $z$ position, the feed rate $f = \dot{x}$ and the number of passes during the flow-forming process (**Figure 13 (b)**). It is possible also adjusting the angular position, and rotational speed of the mandrel and the workpiece by means of the control of the spindle. The machine setup is equipped with force and displacement sensors of the cross support, as well as an encoder to control the angular position of the mandrel.

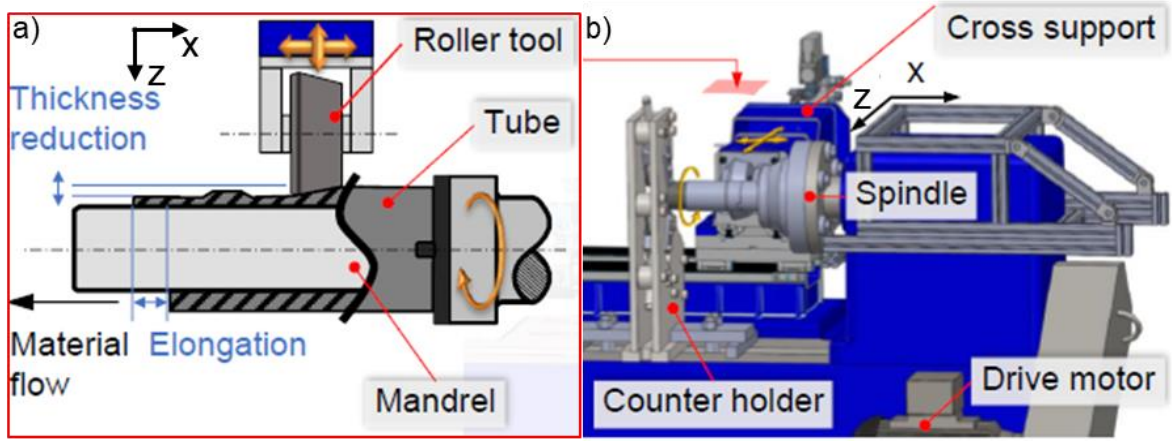

*Figure 13: Reverse flow-forming: (a) Process principle and (b) machine setup.*

To control the martensite content while concurrently ensuring the workpiece geometry, two sensor systems are installed. Regarding the online monitoring of the workpiece dimensions, two laser distance sensors OM70 manufactured by Baumer GmbH (Friedberg, Germany) measure the actual wall thickness reduction. The amount of transformed α'-martensite is monitored by means of a soft sensor composed in its hardware by the micromagnetic testing system 3MAII by Fraunhofer IZFP (Saarbruecken, Germany). In this case, the sensor measures the maximum amplitude of the MBN ($M_{max}$), which has been successfully correlated with the evolution of the strain-induced α'-martensite. The software part of the soft sensor consists of a mathematical model based on empirical data. This model computes the amount of α'-martensite in the workpiece from the micromagnetic measurements under the influence of process parameters like the feed rate. **Figure 14: (a)** illustrates the 3D-surface that represents the mathematical model and its corresponding equation.

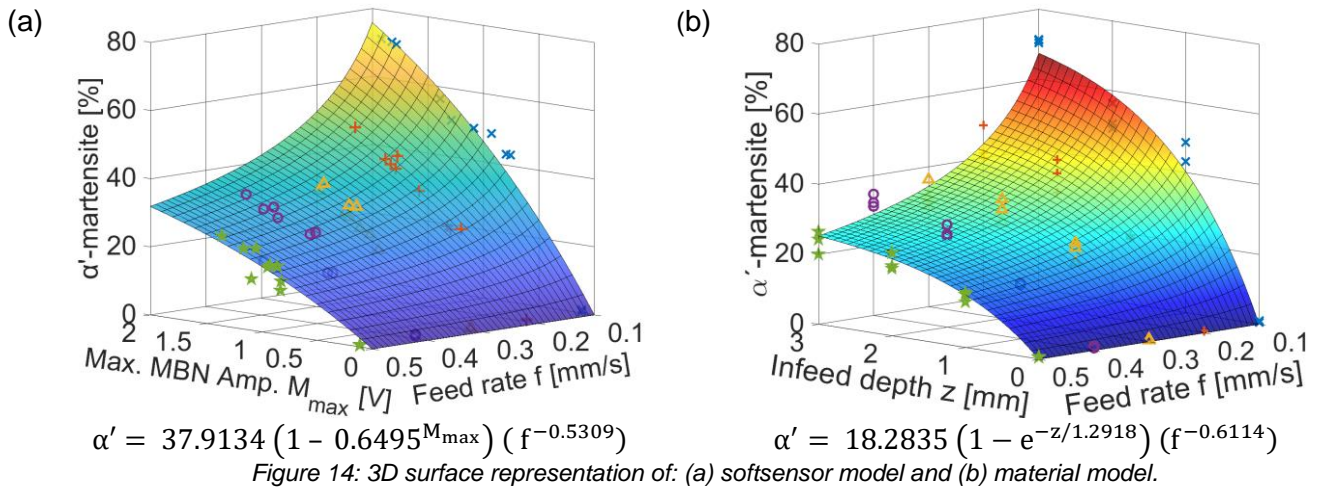

$$\alpha' = 37.9134 \left(1 - 0.6495^{M_{max}}\right) \left(f^{-0.5309}\right)$$

$$\alpha' = 18.2835 \left(1 - e^{-z/1.2918}\right) \left(f^{-0.6114}\right)$$

*Figure 14: 3D surface representation of: (a) softsensor model and (b) material model.*

Under isothermal conditions the variation of process parameters like infeed depth $z$ and feed rate $f$ of the roller tool produces a defined thickness in the final workpiece, which triggers the phase transformation in the microstructure with a determined amount of α'-martensite. To simulate both – geometrical as well as material properties, a control-orientated system model was developed (Kersting et al., 2023). Here, an empirical, multivariate characteristic curve is used as a partial model to calculate the wall thickness resulting from the forming process. This was developed on the basis of the experimental investigations using a polynomial regression approach. For the computation of the martensite content, a material model based on experimental data was developed to calculate the amount of α'-martensite depending on the infeed depth and feed rate of the roller tool (**Figure 14: (b)**). Higher infeed depth and lower feed rate of the roller tool favor the phase transformation process of the material, due to the higher local strain concentrations during flow-forming. This directly describes the coupling between workpiece geometry and the corresponding evolution of properties. The discussed models are applied in real-time as an observer and soft sensor for the property-controlled flow-forming process. Besides online application, they are used to determine the process strategy and to design the control for decoupling geometry and workpiece properties.

In this study case, three different process strategies have been implemented to decoupling the relationship between the workpiece geometry and the evolution of properties. This entails specifically, to achieve a change in product properties, namely the α'-martensite content, without relevant variation of the wall thickness of the workpieces (Arian et al., 2021). The first strategy is carried out under isothermal conditions and consists of an intelligent, coordinated variation of the roller tool feed rate and the infeed depth. It uses the fact that a certain wall thickness reduction could be realized with different parameter combinations that result in different martensite fractions. The second strategy is also performed under isothermal conditions in a multi-stage flow-forming process in which the plastic deformation is carried out by means of several number of passes of the roller tool. The third strategy includes thermomechanical forming in which the α'-martensite formation is locally favored in a single-stage process by means of the use of temperature actuators, namely by means of cryogenic cooling or heat induction (Arian et al., 2023). The application of these strategies enables the production of high-quality components with specific gradation of properties, excellent dimensional accuracy and surface quality.

### 4.7 Freeform bending with movable die (Single-stage, Cold, Ferromagnetic Material)

Freeform bending with movable die is a process allowing the bending of complex geometries without having to change the bending tool. As of now, set geometrical dimensions may be bent that are, meanwhile, tied to a strict set of product properties. However, for the sustainable and economically advantageous process design, a closed-loop control based on the product properties that may separately influence the dimensional accuracy must be established. As a foundation for the successful design of such a closed-loop control, the DOF at disposal in the process need to be combined in such a way that the evolution of product properties are decoupled from the dimensional accuracy (Maier et al., 2021). Product parameters that are especially of interest within this research, are residual stresses, strength as well as the induced plasticity during bending as these particularly dictate further processing steps and service application of the component (Maier et al., 2021; Stebner et al., 2021).

The machine at hand is a 6-axis freeform bending machine, designed by J. Neu GmbH (Grünstadt, Germany). As the name suggests, this machine offers six DOF: translation of the die in two directions on the *xy*-plane, rotation of the bending die around *x*-, *y*- and *z*-direction as well as the movement of the feed along the *z*-direction. When in use, so-called rotation and deflection of the bending die are needed to produce freeform bent parts. In this case, deflection means the translation of the bending die along the *y*- or *x*-axis and rotation along the *x*-, *y*- and *z*-axis. For a better understanding, see **Figure 15 (a)** (Maier et al., 2021). These DOF indicate that the interaction of dimensional accuracy and product properties lies in the superposition of stresses. Perspectively, the forming temperature may present a further DOF for the decoupling of dimensional accuracy and product properties (Maier et al., 2022).

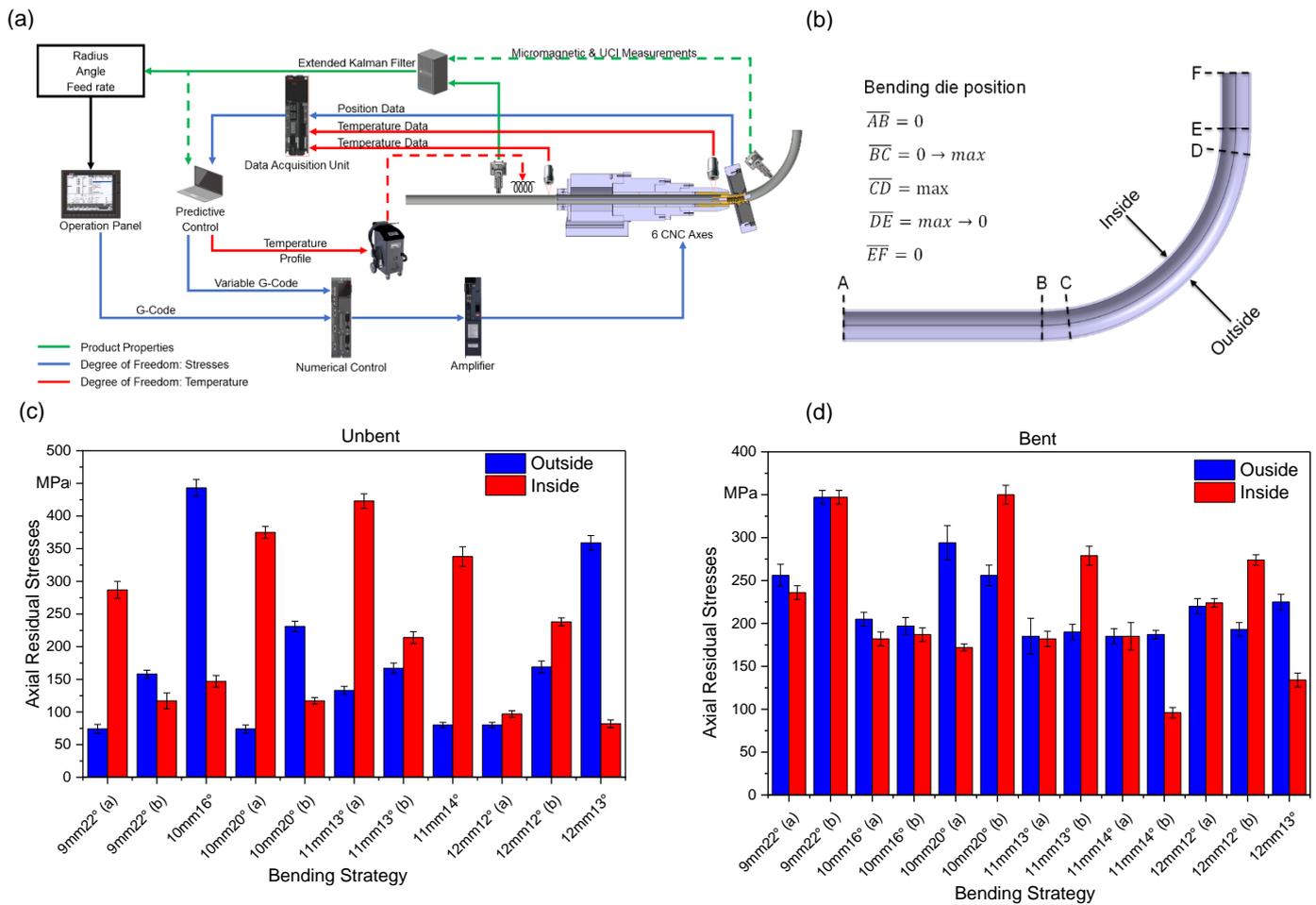

*Figure 15: (a) Schematic depiction of the system of the freeform bending process with movable die and decoupling methods with the respective DOF, (b) path of the bending die position during process, (c) Axial residual stresses in unbent state and (d) in bent state as validation of decoupling product property evolution from the dimensional accuracy.*

The path of the bending die, depicted in **Figure 15 (b)**, is as follows: $\overline{AB}$ is the unbent tube, made up of longitudinally welded tubes with the material P235 TR1, with a distance of 275 mm, followed by the undefined section $\overline{BC}$ where the bending die moves into the maximum position. The maximum position is defined by the communicated final deflection and rotation angle. $\overline{CD}$ is the distance, where only the feed is moving in z-direction and the bending die is held constant at full deflection and angle. In section $\overline{DE}$ the bending die moves back to neutral position and $\overline{EF}$ is again unbent tube.

To influence the evolution of dimensional accuracy and product properties, the authors primarily utilize the superimposition of stresses, in introducing an innovative bending strategy – the non-tangential bending. During this bending strategy, the bending die at final deflection, meaning in section $\overline{CD}$, is not tangential to the tube. Leading to the component being bent slightly over or under due to the bending die being moved to differently combined deflections and rotations (Maier et al., 2021).

Utilizing this novel bending strategy, it is now possible to bend tubes with the same geometry, however independent from product properties namely the axial residual stresses, as can be seen in **Figure 15 (c)** and **(d)**. They show, the axial residual stress state before bending in the steel tubes as well as post undergoing non-tangential bending. It can be seen, that prior to the bending, the axial residual stresses are scattering strongly, while after the bending they can be influenced in a targeted manner as well as decoupled from the dimensional evolution of the tubes, which lays the foundation for a property-based closed-loop control. Now, based upon the understanding of product property evolution and dimensional accuracy, a soft sensor deriving the relevant mechanical properties as well as closed-loop control may be derived, that influences the actuators of the machine. The soft sensor, in this case, relies on two types of sensors: ultrasonic contact impedance (UCI-) and MBN measurements based on which residual stresses, strength and plasticity level predictions can be given under the consideration of measurement and process noise based on a gray-box and state space modeling approach. A closed-loop control can then be designed which will intervene in the actuators position based on the predictions of the soft sensor (Ismail et al., 2021, 2022; Stebner et al., 2021; Stebner et al., 2022).

### 4.8 Press Hardening (Multi-stage, Hot, Ferromagnetic Material)

In press hardening, the thermo-mechanical interactions and their effect on the final product properties are complex and therefore difficult to predict (Neugebauer et al., 2012). For a multi-stage press hardening process, the number of possible interactions increases (Demazel et al., 2018), as well as the number of disturbance variables. With the aim of being able to set the product properties in multi-stage press hardening, here using the example of hardness and thickness distribution, a closed-loop control is utilized allowing for the thermo-mechanical interactions and the disturbance variables to be considered online. As a basis for the application of the control, multi-stage press hardening in a progressive die with the tool layout and actuators presented in **Figure 16 (a)** is assumed. The progressive die is being operated in an industry-grade servo press. Initially, 22MnB5 slit strip with a ferritic-pearlitic microstructure is precut at room temperature into rectangular blanks and a carrier strip. In the first process stage relevant for the process control, the rectangular blanks are heated to the austenitization temperature $T_\gamma$ by means of induction heating. Next in the cooling/heating stage, a temperature profile is set using resistance heating (in area 2) and cooling by compressed air from flat fan nozzles (in area 1).

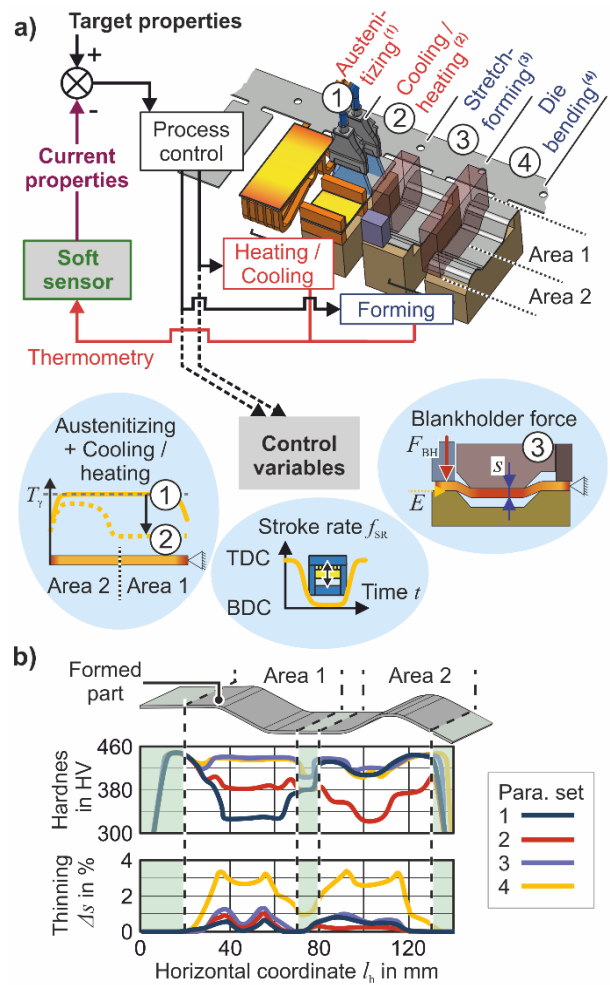

*Figure 16:(a) Process setup and control-loop for multi-stage press hardening in a progressive die;( b) Hardness and thinning distribution of the formed part for different sets of control variables - results from numerical simulation of the multi-stage process.*

In the first forming stage, the blank is stretch-formed into a hat shaped profile. A variable blankholder force $F_{BH}$ enables to adjust the sheet draw-in $E$, hence, the sheet thinning $\Delta s$. During the final process step – the die bending –, one sidewall of the hat shaped profile is bent. In addition, the stroke rate $f_{SR}$ can be altered throughout the process. Depending on the stroke rate, different contact times between the hot sheet and the forming tools are obtained, whereby the effective quenching rate throughout the whole process is varied. For feedback of the current product properties, a soft sensor is employed. The soft sensor is a cascade of three sub-models. First, based on the in-situ temperature measurements with thermocouples in the individual stages and a thermal imaging camera, the temperature distribution is reconstructed as a function of location and time (Kloeser et al., 2021). Then, based on the now known temperature distribution, with the second sub-model, the plasticity in the forming stages and thus the sheet thickness distribution is estimated. The third sub-model then uses the information on the thermal (temperature – 1. sub-model) and mechanical (plasticity – 2. sub-model) history to predict the microstructure evolution and as a result the hardness distribution. Then the soft sensor output is used to adjust the control variables according to the deviation from the target properties.

A multivariable control, in this case the simultaneous control of the sheet thinning $\Delta s$ and the hardness distribution in area 1 as well as area 2 of the formed part (other areas are neglected), can only be accomplished if the target properties can be set in a decoupled manner via the available DOF. To demonstrate how the decoupling can be achieved in the given process with the presented extended actuator setup, the hardness and the sheet thinning distribution of the final formed and quenched part resulting from different parameter sets of control variables listed in **Table 1** are shown in **Figure 16 (b)**. The data is obtained from thermo-mechanical simulation of all process stages with the FEM software LSDyna (solver R12) using the material model Mat248. With the parameter sets 1 to 3, differing hardness distributions in the area 1 and 2 can be produced while maintaining the same sheet thinning (lower than the material thickness fluctuation of ± 2 %). Varying the thermal degress of freedom only, set by the actuators of stage 1 and 2, either minimal hardness in area 1 (para. Set 1) or in area 2 (para. set 2) can be achieved or, alternatively, maximized hardness in both areas (para. set 3). On the other hand, by changing the mechanical parameters as well (para. set 4), controlling the sheet thinning is possible, while maintaining the hardness distribution according to para. set 3 (same thermal control variables). This shows the potential for controlling of various product properties in a decoupled manner.

**Table 1:** Parameter variation of the control variables from Figure 16

| Para. set | Control variable | | | | |
|---|---|---|---|---|---|
| | Stage 1 | Stage 2 | | Stage 3 | Press |
| | $T_\gamma$ in °C | Area 1 cooling | Area 2 (re-)heating | $F_{BH}$ in kN | $f_{SR}$ in n/min |
| 1 | 900 | 16 K/s | to 900 °C | 5 | 6 |
| 2 | 900 | 20 K/s | off | 5 | 6 |
| 3 | 1000 | 20 K/s | to 1000 °C | 5 | 6 |
| 4 | 1000 | 20 K/s | to 1000 °C | 20 | 8 |

## 5. Conclusion and Outlook

In conclusion, this paper offers important insights on decoupling strategies regarding the dimensional accuracy and the product properties of components shaped by different forming processes. This lays an important basis for the implementation of closed-loop property controls within the processes that not only enhance but optimize the processes regarding both sustainable as well as economical aspects in their design.

The following table lists all processes described in **Chapter 4** with the respective actuator/DOF at hand separated into categories of stress and temperature, which are used to influence the dimensional accuracy and mechanical properties. Furthermore, the individual product properties of interest and the investigated materials are listed.

Table 2: Overview of the to be discussed processes, the DOF for influencing the evolution of product property and dimensional accuracy, the to be deduced parameter of interest for inline closed-loop controls as well as the investigated materials.

| **Process** | **DOF in terms of σ** | **DOF in terms of T** | **Parameter of Interest** | **Material** |
|---|---|---|---|---|
| (1) Skin pass-rolling | • Superposition of stresses due to strip tension | | • Surface properties | E-Cu58, DC01 DC04 |
| (2) Flow-Forming | • Feed rate<br>• Infeed | | • Strain hardening | S235+N |
| (3) Open-die forging process | • Forming speed<br>• Degree of deformation<br>• Infeed and Feedrate | • Conductive heating of test specimen<br>• Holding time<br>• Thermal energy due to contact friction | • Microstructure<br>• Grain size (lamellar → globular) | TNM-B1 (Ti-43.5Al-4Nb-1Mo-0.1B) |
| (4) Thermomechanical Tangential Profile Ring Rolling | • Rolling force<br>• Rotational speed | • Temperature rate | • Strength<br>• Surface hardness | 16MnCr5 100Cr6 |
| (5) Punch-hole-rolling | • Infeed<br>• Rotational speed | • Infeed (30 - 100 °C) (holding time/passive) | • Strength<br>• Hardness | 1.4301 1.4404 |
| (6) Reverse Flow-forming | • Feed rate<br>• Infeed<br>• Rotational Speed<br>• Number of passes | • Forming temperature | • α´-martensite fraction<br>• geometry | 1.4307 |
| (7) Freeform Bending | • Feed rate<br>• Bending angle | | • Geometry<br>• Strength<br>• Residual stresses | P235 TR1 |
| (8) Press hardening | • Stroke rate<br>• Blank holder force | • Austenitization temperature<br>• Local temperature distribution<br>• Quenching rate | • Distribution of hardness and thickness | 22MnB5 |

Based on **Table 2**, the following four questions arise:
1. Which actuators are suited best for controlling the target values dimensional accuracy and mechanical properties?

2. Which sensors are useful?
3. Which mathematical models are capable of modeling both geometry and properties?
4. How exactly is the decoupling realized?

The following tables summarize the findings regarding these four research questions, based on the selected choice of forming processes. Furthermore, recommendations regarding actuators in forming processes, sensor choices, mathematical models as well as decoupling strategies are formulated.

Table 3: Summary of which actuators are most effective for a manipulation of mechanical properties and dimensions with respect to forming stages, forming temperature, and investigated material.

| Which actuators are suited best for controlling the target values dimensional accuracy and mechanical properties? | | |
|---|---|---|
| *Forming Stages* | *Forming Temperatures* | *Materials* |
| **Single-Stage Forming:** Actuators using the superposition of stresses are well suited to influence both dimensions and mechanical properties. | **Cold:** In cold forming, the superposition of stresses is the only used DOF. It allows the manipulation of mechanical properties as well as geometry, where in terms of geometry the Bauschinger effect is utilized. | **Ferromagnets:** All actuators can be used to influence the product properties. |
| **Multi-Stage Forming:** The actuator temperature is especially important whenever phases and grain morphology are of importance, while the actuator stress is always suited for a manipulation of the dimensions. | **Hot:** The actuator temperature is always relevant when phases and morphology are important, while stresses set the geometry. | **Paramagnets:** - |
| **Cyclical Forming:** Both actuators temperature and superposition of stresses are important for respective setting of both dimensions and mechanical properties. | | **Meta-stable Paramagnets:** Actuator stress indispensable when TRIP-effect is used. |
| **In Summary:** <br> • **Actuator σ**: Especially relevant in cold forming for the manipulation of both dimensions (exploitation of the Bauschinger effect) and mechanical properties. Dimensions always set via this actuator. <br> • **Actuator T**: Primary actuator in hot forming and always needed when microstructural morphology or phases in steel are relevant product properties. | | |

Table 4: Summary of which sensors are most effective in the detection of a change in product properties with respect to forming stages, forming temperature, and investigated material.

| Which sensors are useful? | | |
|---|---|---|
| *Forming Stages* | *Forming Temperatures* | *Materials* |
| **Single-Stage Forming:** BHN and ECT sensors are very well suited, as they are susceptible to a change in dislocation density and hence, the tied in evolution of product properties. | **Cold:** BHN and ECT sensors are suited, as during cold forming the mechanical properties are largely influenced due to a change in dislocation density, which is very well detected by these types of sensors. | **Ferromagnets:** All types of sensors introduced can be used on ferromagnetic materials. |
| **Multi-Stage Forming:** Temperature sensors always needed when evolution in microstructural morphology and phases are relevant. BHN suited when plasticity induced phase transition is to be monitored. | **Hot:** When grain morphology or phases are parameters of interest, temperature sensors should be used. BHN and ECT sensors are also suitable, however, some sensors are unresistant to heat. Furthermore, for BHN be advised that the magnetic parameters of the material will change when the Curie temperature is reached. | **Paramagnets:** BHN cannot be used on paramagnets. Other sensors can be used. |
| **Cyclical Forming:** Even though small degrees of deformation, BHN and ECT suitable. Again, due to the susceptibility in microstructural changes as grain boundaries and dislocation density. | | **Meta-stable Paramagnets:** BHN can be used, if phase transformation from fcc to bcc (e.g., TRIP steels) occurs. BHN may then be used to correlate phase fractions. |
| **In Summary:**<br>- **BHN sensors:** As the dislocation density is the main influencing factor on the mechanical properties during cold forming, they are very well suited for monitoring cold processes even with little degrees of deformation. Their use is limited to ferromagnetic materials, however, if a meta-stable austenite undergoing the TRIP effect is investigated, BHN may be used for monitoring the induced bcc phase fraction. The authors furthermore advise to be aware of influences due to temperature on the sensors (not equipped to withstand high heat) and the materials (ferromagnetic properties lost at Curie Temperature).<br>- **ECT sensors:** Well-suited for all materials and small degrees of deformation.<br>- **Temperature sensors:** Advised to use if microstructure morphology and phase fractions are important. | | |

Table 5: Summary of which mathematical models are suitable to model the evolution of mechanical properties and dimensions with respect to forming stages, forming temperature, and investigated material.

| **Which mathematical models are capable of modeling both geometry and properties?** | | |
|---|---|---|
| *Forming Stages* | *Forming Temperatures* | *Materials* |
| **Single-Stage Forming:** In single-stage forming, gray-box models in the form of state space observers, such as Extended Kalman Filters, offer a suitable approach for modeling. Black-box models are generally suited for modeling these types of systems, especially when phenomenological knowledge is limited to non-existent. | **Cold:** Gray-box models suited based on state space observers suited. Black-box model convenient, when modeling under uncertainty using e.g., Gaussian Process Regression. | **Ferromagnets, Paramagnets and Meta-stable Paramagnets:** Modeling approach not dependent upon the type of material investigated. |
| **Multi-Stage Forming:** At least gray-box models are necessary. When temperature is a main actuator, numerical simulation of the temperature profile important. When phase fractions influenced by actuators based on stress induction, simple regression possible. | **Hot:** Regressions and characteristic field derivation suited, again categorized as gray-box models.<br><br>When temperature important actuator (grain morphology/phase fractions product properties of interest) simulation models such as Finite Element simulation appropriate. | |
| **Cyclical Forming:** Modeling approach must be gray-box or black-box model. | | |
| **In Summary:** <br> • **White-box models**: White-box models are not suited to fully model the product properties as they develop during the forming process, as there is no complete phenomenological understanding of the influences of the actuators on the product properties. Thus, the modeling approach must be at least based on a gray-box modeling. <br> • **Gray-box models/Black-box models**: Both are suited to model the development of the product properties with respect to the actuators in the process. Black-box models are especially suited when uncertainties are to be modeled. <br> • **Numerical simulation**: As soon as the temperature and morphology of the grains is important, simulation models are important for the modeling of actuators influence on product properties. <br> • **Material** is not a main influencing parameter for choice of modeling approach, rather product properties of interest and actuators in system. | | |

Table 6: Summary on how the decoupling of mechanical properties and dimensions is achieved with respect to forming stages, forming temperature, and investigated material.

| How exactly is the decoupling realized? | | |
|---|---|---|
| *Forming Stages* | *Forming Temperatures* | *Materials* |
| **Single-Stage Forming:** Using superimposition of stresses, non-destructive testing, as well as gray-/black-box models. | **Cold:** Stress as actuator for dimensions and product properties; BHN very well suited as change in product properties during cold forming largely attributed to a change in dislocation density, which can be detected by BHN. Gray-/black-box models recommended. | **Ferromagnets, Paramagnets and Meta-stable Paramagnets:** Both temperature and stress actuators show effect on all types of materials. Sensor choice is only limited in case of paramagnets, as BHN does not work with these types of materials. Mathematical models are not influenced by the investigated material itself, rather by the product properties of interest as well as the actuators utilized. |
| **Multi-Stage Forming:** Temperature for microstructure and phase manipulation, stresses for the dimensions of the workpiece; temperature sensors and magnetic non-destructive testing recommended (however, limited to occurring temperature in the process) and gray-/black-box models. | **Hot:** Temperature as actuator for mechanical properties as function of the microstructure evolution/ phase changes; dimensions set using stresses. Temperature sensors and magnetic testing methods useful (careful as magnetic sensors not all suitable for testing in hot environment), gray-/black-box models supported by simulation of temperature profiles in components. | |
| **Cyclical Forming:** Temperature and stress actuator, BHN and ECT suited especially for limited degrees of deformation and gray-/black-box model. | | |
| **In Summary:** <br> • **Actuators:** Actuators based on temperature and the superimposition of stresses offer strategies to decouple the mechanical properties from the dimensional accuracy. <br> • **Sensors**: Non-destructive testing in the form of BHN and ECT are suited. Especially, during cold forming and low degrees deformation, these types of sensors show very good susception to the change in product properties as a function of the microstructure evolution. Whenever grain morphology or phase fractions are parameters to be controlled, temperature sensors are necessary. | | |

> - **Mathematical Models:** For the mathematical modeling of the processes, at least gray-box modeling is necessary, as there is no full phenomenological knowledge of the systems available. Black-box models also offer a good modeling solution.
> - **Material:** Ferromagnetic materials can be used in all systems with all types of sensors. Paramagnetic materials cannot be investigated using BHN, unless they are meta-stable and convert to bcc structure.

Next to this newly established understanding on how actuators in forming processes influence the mechanical properties and the dimensional accuracy of the component, further research hypotheses arise that will need to be investigated for final implementation of closed-loop property control based on soft sensors:

- As product properties are always a function of the microstructure evolution, characteristic fields between the product properties with regard to the microstructure evolution induced by the forming processes can be derived.
- To establish a real-time closed-loop control, the temporal influences in the system must also be considered. That is, how does the microstructure, ergo the product properties, develop depending on time, at what point in time does the relevant actuator influence the product properties, and finally how quickly can measurements be taken. Regarding the temporal influences will then allow the respective actuator to be controlled precisely at the correct point in time, manipulating the product properties to the desired characteristic.

The authors will work on these topics within the third funding period of the priority program 2183 and will publish their findings.

## **Funding**

This research was funded by the Forschungsgemeinschaft (DFG, German Research Foundation) – 424334318; 424334660; 424334154; 424334856; 424333859; 424337466; 424334423; 424335026.